\documentclass[preprint,showpacs]{revtex4}
\usepackage{epsfig}
\newcommand{\ben}{\begin{eqnarray}}
\newcommand{\een}{\end{eqnarray}}

\newcommand{\bef}{\begin{figure}}
\newcommand{\eef}{\end{figure}}

\begin{document}
\title{Decoupling of the superconducting and magnetic (structural) phase transitions in electron-doped
${\rm BaFe}_2{\rm As}_2$}
\author{P. C. Canfield}
\affiliation{Ames Laboratory and
Department of Physics and Astronomy, Iowa State University, Ames,
IA 50011, USA}

\author{S.~L.~Bud'ko}
\affiliation{Ames Laboratory and
Department of Physics and Astronomy, Iowa State University, Ames,
IA 50011, USA}

\author{Ni Ni}
\affiliation{Ames Laboratory and
Department of Physics and Astronomy, Iowa State University, Ames,
IA 50011, USA}

\author{J.~Q.~Yan}
\affiliation{Ames Laboratory and Department of Physics and
Astronomy, Iowa State University, Ames, IA 50011, USA}

\author{A.~Kracher}
\affiliation{Ames Laboratory and Department of Physics and
Astronomy, Iowa State University, Ames, IA 50011, USA}

\begin{abstract}
Study and comparison of over 30 examples of electron doped ${\rm
BaFe}_2{\rm As}_2$ for transition metal (TM) = Co, Ni, Cu, and (Co/Cu mixtures) have
lead to an understanding that the suppression of the structural/antiferromagnetic
phase transition to low enough temperature in
these compounds is a \emph{necessary} condition for
superconductivity, but not a \emph{sufficient} one. Whereas the
structural/antiferromagnetic transitions are suppressed by the
number of TM dopant ions (or changes in the $c$-axis) the
superconducting dome exists over a limited range of values of the
number of electrons added by doping (or values of the ${a/c}$ ratio).
By choosing which combination of dopants are used we can change
the relative positions of the upper phase lines and the
superconducting dome, even to the extreme limit of suppressing the
upper structural and magnetic phase transitions without the
stabilization of low temperature superconducting dome.

\end{abstract}
\date{\today}
\maketitle

The discovery of superconductivity in the LaFeAsO [1] and ${\rm
BaFe}_2{\rm As}_2$ [2] systems has lead to a renaissance in
interest in transition metal based superconductivity.
Both of these systems manifest substantial $T_c$ values when the
structural/antiferromagnetic phase transitions are sufficiently
suppressed by substitution on the alkali-earth, transition metal
and/or oxygen site. Although the systematic studies of F- and
K-doping have been difficult due to problems in controlling and
assessing stoichiometry, transition metal doping, especially of
the ${\rm BaFe}_2{\rm As}_2$ system has been tractable and quantifiable.

In the case of ${\rm Ba(Fe}_{1-x}{\rm Co}_x)_2{\rm As}_2$  a
comprehensive, and highly reproducible, $T(x)$ phase diagram has
been determined [3] and confirmed/reproduced by several groups
[4-6]. The structural phase transition is suppressed by roughly
15 K per atomic percent Co and increasingly separates from the
lower, magnetic phase transition as more Co is added [3,4,7,8]. For
intermediate doping levels, superconductivity has been observed to
strongly interact with the magnetic order and fluctuations in the
antiferromagnetically ordered, orthorhombic state [7]. For higher
Co doping levels both the structural and antiferromagnetic phase
transitions are suppressed and superconductivity occurs in the
tetragonal phase. These data are all consistent with the idea that
superconductivity is stabilized when the tetragonal phase is
brought to "low enough" temperatures by perturbing the parent
compound. This may be associated with reducing the size of the
orthorhombic distortion and ordered moment "enough" or bringing
the magnetic fluctuations associated with the tetragonal phase to
"low enough" temperatures. Superconductivity does not require the
complete suppression of the orthorhombic/antiferromagnetic
phase, just its suppression to an adequately low temperature
[3-6].

There is a clear correlation between the upper (structural
and magnetic) phase transitions and the lower temperature,
superconducting phase, but, to date, it is a qualitative one at
best. In this Letter we have studied over 30 samples of electron
doped ${\rm BaFe}_2{\rm As}_2$ where the electron doping is coming
from 3$d$ transition metal substitutions on the Fe site. We
have grown and examined single crystalline samples of the ${\rm
Ba(Fe}_{1-x}{\rm TM}_x)_2{\rm As}_2$ system for TM = Co, Ni, Cu,
and (Co/Cu mixtures) and find that whereas the suppression of the
upper structural phase transitions is a \emph{necessary} condition
for low temperature superconductivity, it is not a
\emph{sufficient} one. This distinction can be understood by our
observation that whereas the upper transitions appears to be
suppressed by the number of impurity atoms substituted for Fe (or
the change in the crystallographic $c$-axis) the location and extent
of the superconducting dome scales with the number of additional
electrons, one for each Co, two for each Ni and three for each Cu
atom (or the change in the ratio or the crystallographic $a$-axis to
$c$-axis). By choosing which combination of dopants are used, we can
change the relative positions of the upper phase lines and the
superconducting dome, even to the extreme limit of suppressing the
upper structural and magnetic phase transitions without the
stabilization of low temperature superconducting dome.

Single crystals of ${\rm Ba(Fe}_{1-x}{\rm TM}_x)_2{\rm As}_2$
system for TM = Ni, Cu, and (Co/Cu mixtures) were grown in a
similar manner as the ${\rm Ba(Fe}_{1-x}{\rm Co}_x)_2{\rm As}_2$
compounds [3]. Actual doping levels (rather than nominal) were
determined via WDS analysis using an electron probe microanalyzer
of a JEOL JXA-8200 electron microprobe and are denoted as ${x_{WDS}}$.
Powder X-ray diffraction spectra with Si standard were measured using a Rigaku MiniFlex
and unit cell parameters were extracted using "UNITCELL" analysis
package. Although we attempted to synthesize similar doping levels
of the various Co, Ni, Cu and Co/Cu series by using identical
nominal values, experimentally determined doping
levels revealed slightly different actual values of incorporation
of these different TM dopants. Electrical resistivity measurements
were made using a standard 4-probe configuration and Quantum
Design PPMS (Physical Property Measurement System) and MPMS
(Magnetic Property Measurement System) units to provide the
temperature/field environment. Although single crystals can be
shaped into well defined geometries, the ${\rm AEFe}_2{\rm As}_2$
materials are prone to exfoliation along the $c$-axis that can lead
to spurious resistivity values due to poorly defined current path
lengths and cross-sections \cite{Ni, tanatar1, tanatar2}. For this reason normalized
resistivity values are plotted. Although only resistivity data is
presented in this Letter, detailed magnetization and specific heat
data have also been collected; as in the case of ${\rm
Ba(Fe}_{1-x}{\rm Co}_x)_2{\rm As}_2$ [3], these thermodynamic data
further support the $T(x)$ phase diagrams we infer from transport data.

Figures 1a and 1b present the temperature dependent, normalized
resistivity for ${\rm Ba(Fe}_{1-x}{\rm TM}_x)_2{\rm As}_2$ system
for TM = Co and Ni respectively. For each TM dopant there is a
clear suppression (and separation) of the upper transitions with
increasing $x$ and superconductivity is clearly stabilized once the
structural/magnetic phase transitions are sufficiently
suppressed and exists in both the orthorhombic/antiferromagnetic
phase as well as in the tetragonal one at high dopings \cite{Ni, Dan, Nidope}. Although
${\rm BaCu}_2{\rm As}_2$ itself appears to be a
relatively innocuous compound \cite{Cu, Cutheory}, the ${\rm Ba(Fe}_{1-x}{\rm
Cu}_x)_2{\rm As}_2$ series (Fig. 1c) reveals a key difference:
although the signature of the structural/antiferromagnetic phase
transition is suppressed in a manner similar to that seen for TM =
Co and Ni, there is no superconductivity found for any $x$ value
tried (up to values six times greater than the $x$ = 0.061 shown).
This means that the signatures of the orthorhombic/antiferromagnetic
transitions are not truncated by superconductivity and can be observed
to fade as $x$ is increased.

In order to clarify the effect of Cu as a dopant (i.e. Is it
particularly pernicious to superconductivity or is it essentially
part of a continuum that contains Co and Ni dopants as well?) we
studied a ${\rm Ba(Fe}_{1-x-y}{\rm Co}_x{\rm Cu}_y)_2{\rm As}_2$
series ($x\sim 0.022$ and $0\leq y < 0.05$).  Fig. 1d presents
selected normalized resistivity plots for this series.  As can be
seen in Fig. 1a, a Co-doping of $x$ = 0.024 is insufficient to
induce superconductivity, but additional doping by Cu (Fig. 1d)
can indeed induce superconductivity.  These data clearly show that
Cu is not inherently antithetical to the superconducting state and
that there may well be a deeper and more profound realization to
be made based on these data.

The data presented in Fig. 1 can be summarized in a ${T-x}$ phase
diagram.  The transition temperature values for the upper
structural and magnetic phase transitions were inferred from these
data in manner similar to that used in reference 3 and
subsequently supported by microscopic measurements \cite{Dan, Lester}.  For the
higher Cu concentrations ($x$ = 0.05 and 0.061) the resistive
features

become so broad that the error bars associated with the
determination of the upper (only detectable) transition are
defined by the temperature of the resistance minima on the high
side and the temperature of the inflection point on the low side.
Fig. 2a displays the ${T - x}$ phase diagram for each of the ${\rm
Ba(Fe}_{1-x}{\rm TM}_x)_2{\rm As}_2$ (TM = Co, Ni, Cu, and Co/Cu)
series. Whereas the suppression of the upper phase transitions for
each of these different series appear to depend on $x$ in a similar
manner, the occurrence of superconductivity is not well described
by this parameterization. Superconductivity is found for an
wide range of Co doping values, a narrower range of
Ni doping values and a even narrower range of Cu doping values (in
the ${\rm Ba(Fe}_{1-x-y}{\rm Co}_x{\rm Cu}_y)_2{\rm As}_2$
series).

There is, of course, a second way of plotting these data:
transition temperature as a function of extra conduction electrons
added by the dopant, i.e. grossly assuming the validity of a rigid
band approximation for these dopants.  For TM = Co, the number of
impurity atoms, $x$, per TM site is the same as the number of extra
electrons, $e$, per TM site. When TM = Ni or Cu, this is not the case. A
second parameterization of the data inferred from Fig. 1 is show
in Fig. 1b: a ${T - e}$ phase diagram, where $e$ is the number of extra
electron added per Fe/TM site (for the case of Co ${e = x}$, for the
case of Ni ${e = 2x}$, for the case of Cu ${e = 3x}$). This
parameterization does a much better job of unifying the
superconducting domes of these compounds, but clearly does a much
poorer job of capturing the physics of the suppression of the
upper structural/antiferromagnetic phase transitions.

Although $x$ and $e$ are intuitive (and relatively easy to determine)
parameters, they are certainly not unique ones. Figs. 3a-d
demonstrate that whereas the $c$-lattice parameter variation is
similar to $x$, the variations in the $a$-lattice parameter, the
volume, and the ${a/c}$ ratio do not show universal behavior when
plotted as a function of $x$. This means that the statement that
the upper structural and antiferromagnetic phase transitions scale
with $x$ is equivalent (experimentally) with the statement that they
scale with the variation in the $c$-lattice parameter.

Further examination of Figs. 3a-d reveals that whereas a change of
parameter from $x$ to $e$ will not lead to a collapse of the data for
$a/a_0$, $c/c_0$ or $V/V_0$ onto a universal curve, the variation
of the ${a/c}$ data appears promising, showing variations with $x$ that
differ by factors of two and three. Fig. 3e plots the variation
of $a/c$ as a function of $e$. As clearly shown, ${a/c}$ and $e$ are
experimentally equivalent variables (for 3$d$ TM electron doping) as
well.

One obvious parameter that has not been examined in this study is
the As-Fe-As bonding angle. Unfortunately this was not extracted
from our diffraction data, and given that the location of the As
site is free to vary, it is hard to model. Future measurements
will have to determine whether this angle is related to either $x$
or $e$.

The phase diagrams in Fig. 2 provide graphic evidence that the
structural/antiferromagnetic phase transitions and the occurrence
of superconductivity depend on different parameters for electron
doping via TM substitution: number of impurities (change in
$c$-axis parameter) and number of additional electrons (${a/c}$ ratio)
respectively. This difference allows for the decoupling of these
transitions and the ability to realize that the suppression of the
structural/antiferromagnetic phase transition to low enough
temperature is a \emph{necessary} condition for superconductivity,
but not a \emph{sufficient} one. The data from the ${\rm
Ba(Fe}_{1-x}{\rm Cu}_x)_2{\rm As}_2$ series clearly demonstrate
that if too many electrons are added in the process of suppressing
the structural/antiferromagnetic phase transition the
superconducting dome can be overshot, i.e. by the time the
structural/antiferromagnetic transition is suppressed enough,
too many conduction electrons have been added and window for
superconductivity has been missed. A closer examination of Fig.
2b brings this point even further into focus: although the
superconducting dome for the ${\rm Ba(Fe}_{1-x}{\rm Co}_x)_2{\rm
As}_2$ , ${\rm Ba(Fe}_{1-x}{\rm Ni}_x)_2{\rm As}_2$, and ${\rm
Ba(Fe}_{1-x-y}{\rm Co}_x{\rm Cu}_y)_2{\rm As}_2$ series are
essentially indistinguishable on the higher doping side, they
differ, somewhat, on the lower doping side. This difference would
be consistent with needing to bring the upper transition to low
enough temperature to allow the superconductivity to turn on:
${\rm Ba(Fe}_{1-x}{\rm Co}_x)_2{\rm As}_2$ with its more rapidly
decreasing upper transitions manifest superconductivity at
slightly earlier e-values than the Ni-doped or Cu/Co doped series.

The observation that the upper transitions depend on either the
number of TM dopant atoms added, $x$, or, equivalently, the change
in the $c$-axis dimension, leads to two differing scenarios for what
physical parameter controls this suppression. If $x$ is the salient
parameter, then the upper transitions are controlled by local
physics such as vacancies on the Fe sublattice or the disruption
of very short range fluctuations. On the other hand if the size
of the $c$-axis parameter is the salient variable, then details of
band structure (nesting or not) or degree of As-As bonding across the
Fe-plane would be more likely to control/affect the value of the
upper transition temperatures.

The observation that the superconducting dome is delineated by a
minimum and maximum number of extra conduction electrons (or
possibly ${a/c}$ ratio) provides a clear theoretical constraint/test
for current theories of superconductivity in these fascinating,
complex and potentially useful \cite{theory} compounds.

In conclusion, the study and comparison of over 30 examples of
electron doped ${\rm Ba(Fe}_{1-x}{\rm TM}_x)_2{\rm As}_2$ have lead
to an understanding that the suppression of the structural/antiferromagnetic
phase transition to low enough temperature in
these compounds is a \emph{necessary} condition for
superconductivity, but not a \emph{sufficient} one. Whereas the
structural/antiferromagnetic transitions are suppressed by the
number of TM dopant ions (or changes in the $c$-axis) the
superconducting dome exists over a limited range of values of e,
the number of electrons added by doping (or values of the ${a/c}$
ratio). As clearly shown by the ${\rm Ba(Fe}_{1-x}{\rm
Cu}_x)_2{\rm As}_2$ series, if too many electrons are added per TM
dopant, then the window for superconductivity can be completely
missed. Further work, including the quantitative and comparative analysis of
K-doping and TM-based hole doping, as well as 4$d$ and even 5$d$
TM-based electron doping will have to be carried out to see how
general this decoupling of the structural and superconducting
transitions is and perhaps help resolve which parameterization it
the physically most relevant.

Acknowledgements: We would like to thank N.H.Sung for help in the samples growth. Work at the Ames Laboratory was supported by
the Department of Energy, Basic Energy Sciences under Contract No.
DE-AC02-07CH11358.


\clearpage

\bef \psfig{file=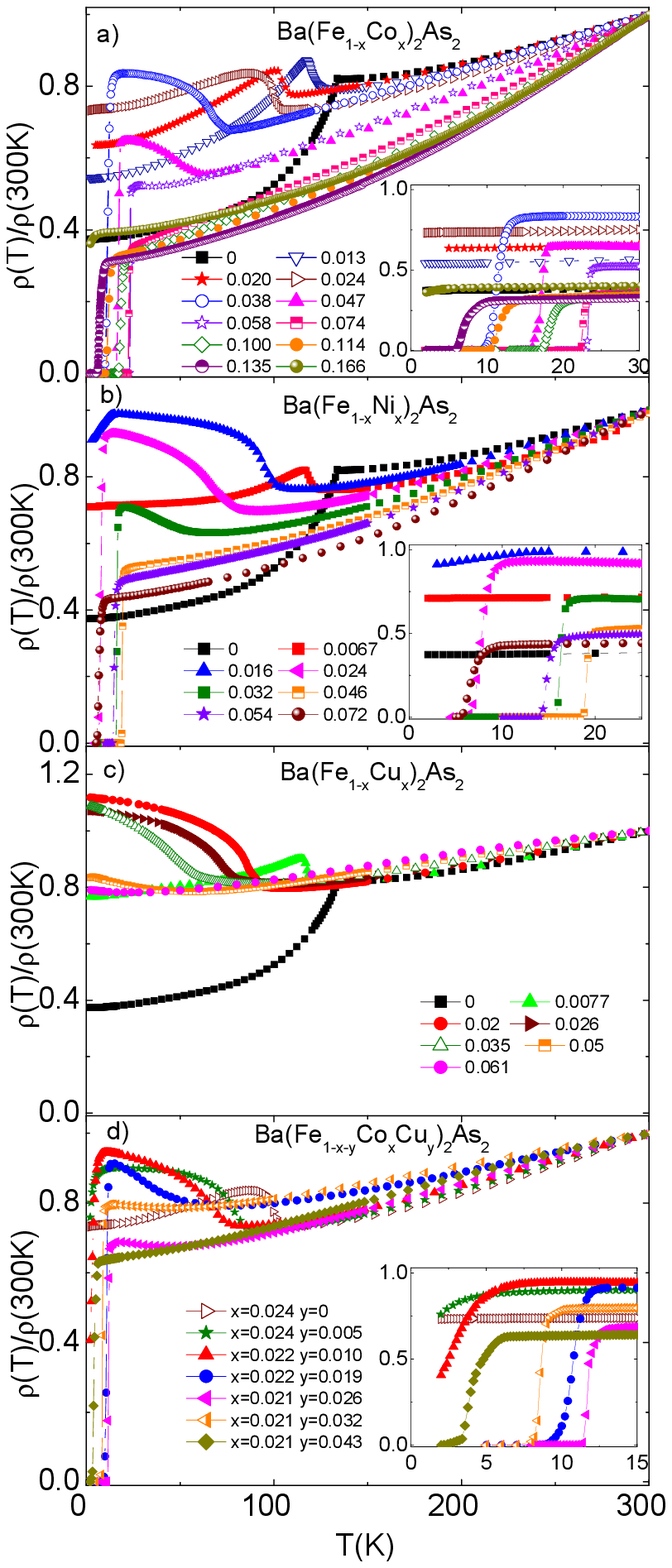,width=3in} \caption{The temperature
dependent resistivity, normalized by room temperature value, for
electron doped ${\rm Ba(Fe}_{1-x}{\rm TM}_x)_2{\rm As}_2$  (TM =
Co, Ni, Cu, and Co/Cu) series:  (a) ${\rm Ba(Fe}_{1-x}{\rm
Co}_x)_2{\rm As}_2$ [3]. Inset: low temperature data for ${\rm Ba(Fe}_{1-x}{\rm
Co}_x)_2{\rm As}_2$ (b) ${\rm Ba(Fe}_{1-x}{\rm Ni}_x)_2{\rm
As}_2$. Inset: low temperature data for ${\rm Ba(Fe}_{1-x}{\rm Ni}_x)_2{\rm
As}_2$ (c) ${\rm Ba(Fe}_{1-x}{\rm Cu}_x)_2{\rm As}_2$ (d)
${\rm Ba(Fe}_{1-x-y}{\rm Co}_x{\rm Cu}_y)_2{\rm As}_2$.
Inset: low temperature data for ${\rm Ba(Fe}_{1-x-y}{\rm Co}_x{\rm Cu}_y)_2{\rm As}_2$}
\label{fig4} \eef

\clearpage

\bef \psfig{file=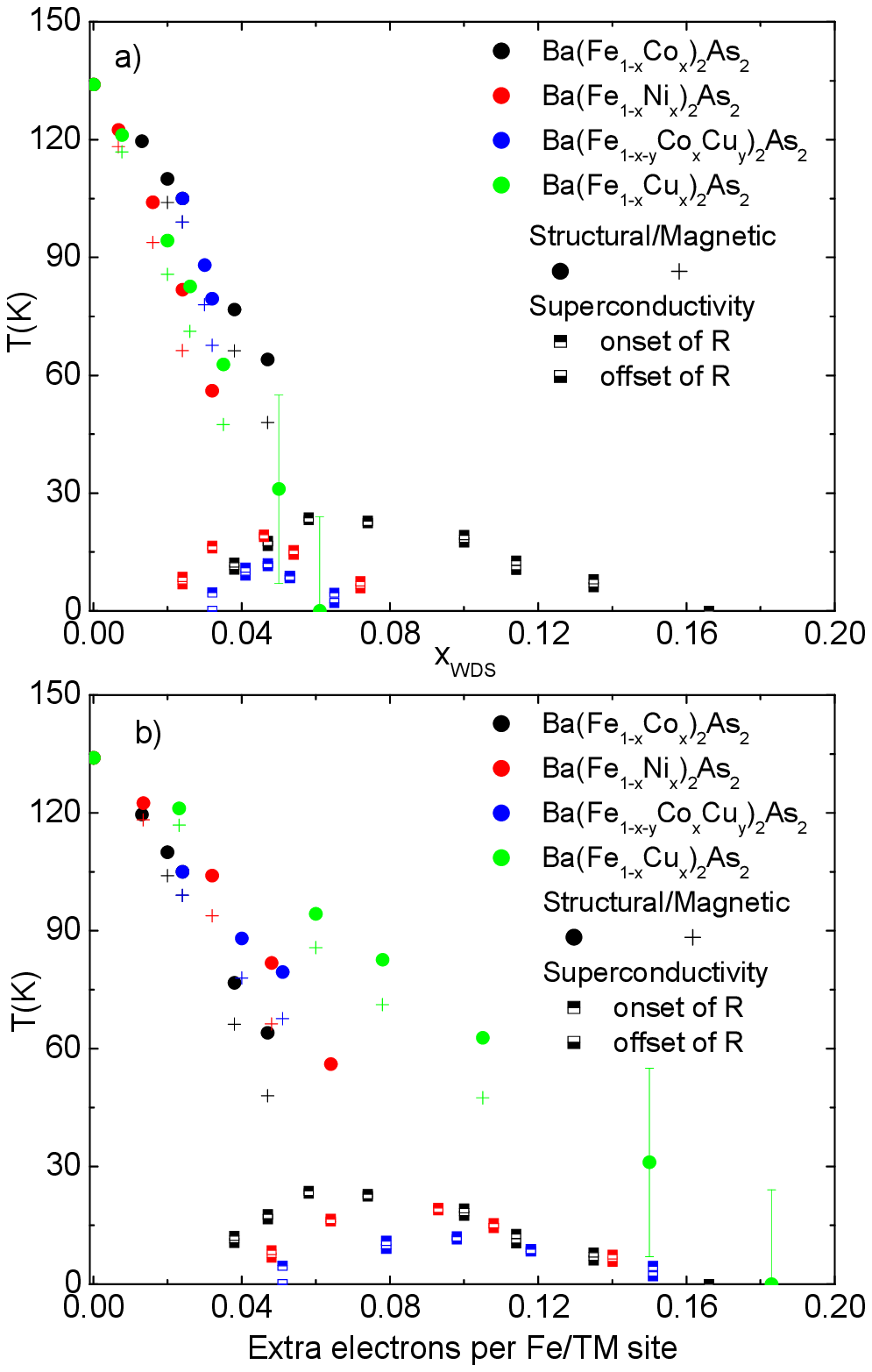,width=3in} \caption{(a) Transition
temperature as a function of the number of substitutional
transition metal ions per Fe site; (b) Transition temperature as a
function of extra electrons contributed by TM substitution per Fe
site. For both plots the transition temperatures were determined
in a manner similar to that described in [3] and the text.}
\label{fig2} \eef

\clearpage

\bef \psfig{file=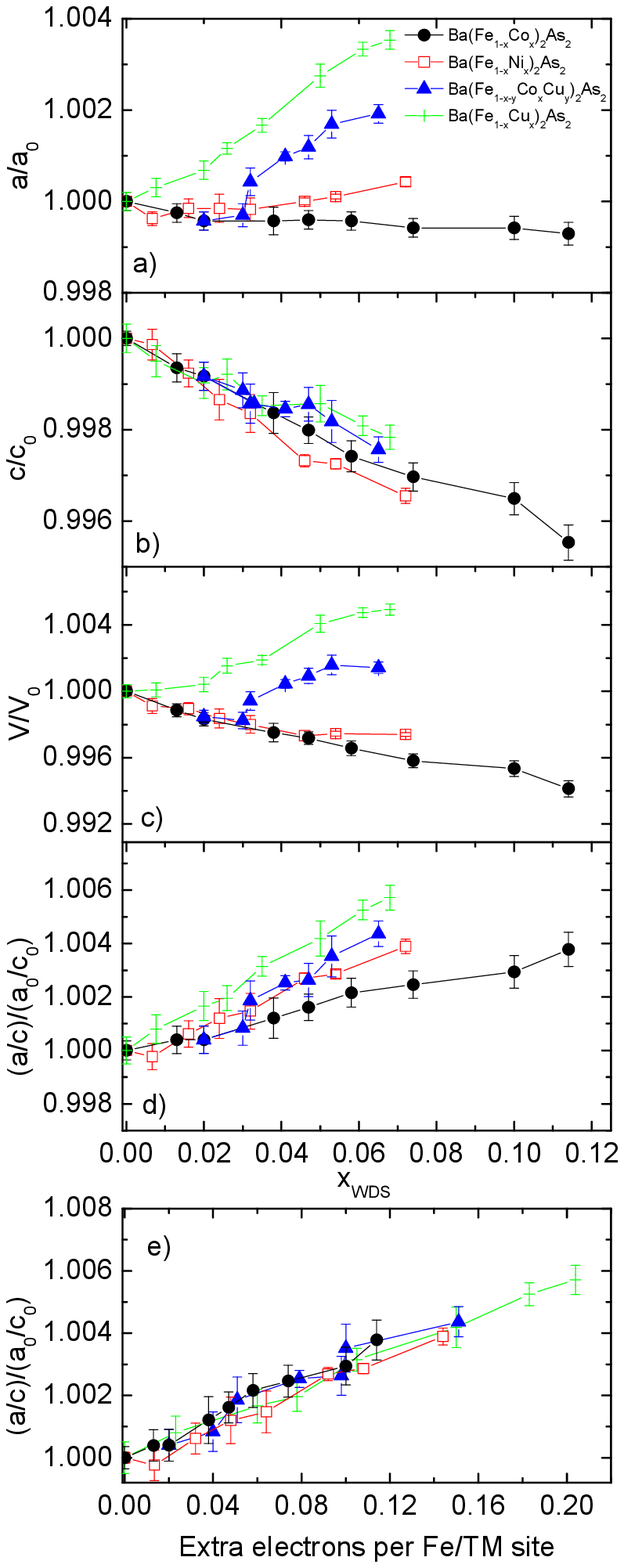,width=3in} \caption{Normalized
structural parameters measured at $\sim$ 300K. (a) $a/a_0$, (b) $c/c_0$, (c) $V/V_0$ and
(d) $(a/c)/(a_0/c_0)$ as a function of transition metal doping,
$x$, and (e) $(a/c)/(a_0/c_0)$ as a function of extra conduction
electrons, $e$. ($a_0$=3.9621$\AA$, $c_0$=13.0178$\AA$)} \label{fig4} \eef

\end{document}